\newcount\opcion
\newlength{\anchopar}
\def \bfr {\begin{flushright}}
\def \efr {\end{flushright}}
\def \caja {\makebox[3.2cm][1]}
\def \v {\vskip}

\def \L {{\cal L}}
\def \R {{\hbox{\it R}}}
\def \d {\hbox{d}\,}
\def \square {\hbox{$\sqcup\!\!\!\!\sqcap$}}

\def \e {\hbox{e}}

\def \p {\partial}

\def \pp {\partial_+}
\def \mm {\partial_-}

\def \ba {\begin{array}}
\def \ea {\end{array}}

\def \bea {\begin{eqnarray}}
\def \eea {\end{eqnarray}}

\def \be {\begin{equation}}
\def \ee {\end{equation}}

\def \bfr {\begin{flushright}}
\def \efr {\end{flushright}}
\def \caja {\makebox[3.2cm][1]}
\def \l {\lambda}
\def \vp {\varphi}
\def \hi {\hbox{i}}
\def \g {{\widehat g}}
\def \hG {{\widehat\Gamma}}
\def \n {{\widehat\nabla}}

\def \spacelike {\>\>{}^{\hbox{\tiny{spacelike}}}
\!\!\!\!\!\!\!\!\!\!\!\!\!\!{}\longrightarrow\>\>}

\def \sspacelike {\>\>\>{}^{\hbox{\tiny{spacelike}}}
\!\!\!\!\!\!\!\!\!\!\!{}\sim\>\>\>\>}

\documentstyle[a4,psfig]{article}

\ifnum\opcion=1
\else

\fi

\begin{document}


\pagestyle{empty}
\bfr
\caja{{\tt FTUV/94-65}}\\
\caja{{\tt IFIC/94-62}}\\
\caja{{\tt Imperial-TP/93-94/60}}\\
\caja{{\tt hep-th/9411105}}
\efr

\vfill

\begin{center}

{\bf\Large Diffeomorphisms, Noether Charges and Canonical Formalism
	in 2D Dilaton Gravity}
\footnote[2]{Work
partially supported by the C.I.C.Y.T. and the D.G.I.C.Y.T.}

\v .5cm

Jos\'e Navarro-Salas$^{1,2}$, Miguel Navarro$^{3,}$\footnote{On leave of
absence from [2] and {\it Instituto Carlos I de F\'\i sica Te\'orica y
Computacional, CSIC, and Facultad  de  Ciencias, Universidad de Granada,
Campus de Fuentenueva, 18002, Granada, Spain.}}\\
and C\'esar F. Talavera$^{1,2}$
\end{center}

\v .5cm

\begin{enumerate}

\item Departamento  de  F\'\i sica  Te\'orica, Burjassot-46100,
Valencia, Spain.

\item IFIC, Centro Mixto Universidad de
Valencia-CSIC, Burjassot-46100, Valencia, Spain.

\item The Blackett Laboratory, Imperial College, London SW7 2BZ,
United Kingdom.

\end{enumerate}

\vfill

\centerline{\bf Abstract}

	We carry out a parallel study of the covariant phase space and
the conservation laws of local symmetries in two-dimensional dilaton gravity.
Our analysis is based on the fact that the Lagrangian can be
brought to a form that vanishes on-shell giving rise to a well-defined
covariant potential for the symplectic current. We explicitly compute
the symplectic structure and its potential
and show that the requirement to be finite and independent of the
Cauchy surface restricts the asymptotic symmetries.

\v 1cm

\noindent PACS number(s): 04.20.Cv, 04.20.Fy, 12.25.+e

\vfill
\eject

\setcounter{page}{1}
\pagestyle{plain}

\section{Introduction.}

        Great efforts have been developed recently on the study of
two-dimensional dilaton gravity theories.
The reason for this interest is that these theories serve as toy
models in which we can develop and test techniques and methods
to be further applied to more realistic (higher dimensional) gravity
theories.
Remarkably, the string inspired model (CGHS-model) of Ref. \cite{CGHS}
(see also \cite{Witten}) admits black hole solutions and, therefore,
provides an interesting toy model to study black hole issues.

        One of the aims of this paper is to study the reduced phase
space of the CGHS-model. This point could be of great interest for
a non-perturbative canonical quantization of the theory. Our
work is based on the covariant phase-space formalism
\cite{Crnkovic}-\cite{Nosotrosii}
and extends the results of a previous paper \cite{Navarro}.
The covariant formalism has already been applied to the CGHS-model
in Ref. \cite{Mikovic,Soh} although their results are valid
for the case of a closed space only.

	Moreover, for Lagrangians vanishing
on-shell, the Noether's procedure can be incorporated, in a
rather natural way, to the covariant canonical formalism.
Therefore, we shall also study, in a parallel way,
the covariant phase space and the conservation laws associated with
diffeomorphism invariance. Our analysis will shed new insight on the
controversy about the notion of mass in 2D dilaton gravity
(see \cite{Bilal}-\cite{Bak}).

	In Section 2 we present briefly the covariant phase space
formalism pointing out the fact that, for Lagrangians vanishing
on-shell, the space of solutions can be endowed with a natural
potential for the symplectic structure. The Noether
charge technique is naturally incorporated in this scheme.
In Section 3 we study in a systematic way the conservation laws
associated with the diffeomorphism invariance and, in particular,
with the asymptotic (Poincar\'e) symmetries of the CGHS model.
In Section 4 we determine the symplectic potential of
the CGHS model. The condition of having a well-defined
potential (i.e. finite and independent of the Cauchy surface)
will restrict the allowed asymptotic symmetries. The Lorentz
symmetry break down and the spatial translation
turns out to be a gauge-type transformation. This will
permit to understand the results of Section 3.
We shall also consider, in Section 5, the case of
spherically symmetric $3+1$ Einstein gravity,
which can also be regarded as a 2D dilaton gravity model
(see \cite{Kuchar} for a related perspective).
Although the stringy and Schwarzschild black holes have the
same canonical structure they differ in the
form of the potential. As a byproduct, this accounts for
the numerical factors in the Komar-type formulas for the
mass in gravity models. We state our conclusions in
Section 6.

\section{Covariant phase space and conservation laws}

	Given a field theory with dynamical fields $\Psi^\alpha(x)$
and action $S=S(\Psi^\alpha(x))$, the phase space can be
defined, in a covariant way, as the space of solutions of the
classical equations of motion. The standard formula
\be
\delta S(X^\alpha) = \int_{\cal M} {\delta S \over \delta \Psi^\alpha}
	X^\alpha + \partial_\mu j^\mu(\Psi^\alpha, X^\beta)
\label{dosi}
\ee
can be interpreted now as the exterior derivative of $S$, on the
covariant phase space, acting on a tangent vector $X^\alpha$
(which solves the linearized equations of motion). In contrast
with the variational calculus which takes the variation $X^\alpha$
vanishing on the boundary of $\cal M$, it is now the first term of the r.h.s.
of (\ref{dosi}) which vanishes automatically. Therefore, the covariant
phase space can be equipped with a presymplectic two-form
\be
\omega = \int_\Sigma \delta j^\mu d\sigma_\mu \, ,
\label{dosii}
\ee
where $\Sigma$ is a Cauchy hypersurface and $\delta$ stands for
the exterior derivative operator. Due to the fact that the symplectic current
$\omega^\mu = \delta j^\mu$ is conserved, the presymplectic form
(\ref{dosii}) is, in general grounds, independent of the Cauchy surface
with a suitable choice of boundary conditions.

	From the above expression it is clear that the one-form
\be
\theta = \int_\Sigma j^\mu d\sigma_\mu
\label{dosiii}
\ee
could serve as a potential for the presymplectic form (\ref{dosii}).
However, $j^\mu$ is not, in general, conserved and hence $\theta$ is
not well-defined.

	Now, let us suppose that the presymplectic
potential current $j^\alpha$ is itself conserved,
\be
\p_\alpha j^\alpha_{|_{sol}}=0 \, .
\label{c1}
\ee
Then, for any field $X\sim \delta \Psi^a$ satisfying the linearized
equations of motion, we will have that
$J^\alpha_X=\hi_X j^\alpha$ is a conserved current:
\be
\p_\alpha J^\alpha_X{_{|_{sol}}}=0 \, .
\label{c2}
\ee

	What is the condition for a presymplectic potential current
to be conserved? On solutions we have
\be
\p_\alpha j^\alpha_{|_{sol}} = \delta {\cal L}_{|_{sol}}\, .
\label{c3}
\ee
Therefore, it is enough that the Lagrangian vanishes on the covariant
phase space.
In this situation the one-form (\ref{dosiii}) is well defined (with
appropriate boundary conditions), $J^\alpha_X = j^\alpha(X)$
coincides with the Noether current and $\theta(X)$ is the corresponding
Noether charge.

\section{Energy-momentum conservation in the CGHS model.}
\label{a}

	The action of the CGHS model is:
\bea
S_{CGHS} &=& \frac{1}{2}\int_{\cal M}\,d^2x\sqrt{-g}
\left[\e^{-2\Phi}(R+4(\nabla\Phi)^2 +4\lambda^2)-
\frac12(\nabla \phi_i)^2\right]
\> .
\label{CGHS1}
\eea
By doing $\varphi=\e^{-\Phi}$ we obtain
\bea
S_{CGHS} &=& \frac{1}{2}\int_{\cal M}\,d^2x\sqrt{-g}
\left[(R\varphi^2+4(\nabla\varphi)^2 +4\lambda^2\varphi^2)-
\frac12(\nabla \phi_i)^2\right]
\>,
\label{CGHS2}
\eea
which, for our purposes, is a form of the action more easy to deal with.

	Now, it is convenient to define a new metric $\g_{\nu\mu}$ by means of
\be
g=\varphi^{-2}\g \, ,
\label{E2}
\ee
in term of which the action takes a remarkably simpler form:
\be
S_{CGHS}=\frac12\int_{\cal M} \d^2
x\sqrt{-\g}\left[\left(\hat R\varphi^2+4\l^2\right)-
\frac12(\nabla \phi_i)^2\right]\, ,
\label{CGHS3}
\ee
The new variable $\widehat g$, which allows to eliminate the kinetic
term in the action, also
emerges in the gauge-theoretical formulation \cite{Cangemi} of the theory,
and in more general models \cite{Louis}.

	The equations of motions are given by:
\bea
\widehat R=0\> &,&
	\quad \widehat{\square}\vp^2=4\l^2
	\quad , \quad \widehat{\square}\phi_i = 0
	\>,
\label{hatEM}\label{mov1}\\
\n_\mu\n_\nu\vp^2 &=& \frac{1}{2} \widehat{\square}\vp^2 +
	\frac12\left(\frac12(\n\phi_i)^2\right)\g_{\mu\nu}-
	\frac12\n_\mu \phi_i\n_\nu \phi_i \> ,
\label{mov2}
\eea
and, if we add a convenient total divergence to the
action of the CGHS model in (\ref{CGHS3}) we can easily bring it to
a form vanishing on-shell
\be
\widehat S_{CGHS}=\int_{\cal M}\widehat\L_{CGHS}
= \frac12\int_{\cal M} \d^2x\sqrt{-\g}\left[\left(\hat R\varphi^2
+ 4\l^2-\n_\alpha\n^\alpha\varphi^2\right)+
\frac{1}{2} \phi_i \widehat{\square} \phi_i\right]
\label{CGHS4}
\>.
\ee

The symplectic potential associated to the above Lagrangian is:
\bea
\hat j^\alpha
&=&\frac12\sqrt{-\g}\left[-
\varphi^2\left(\g^{\mu\nu}\n^\alpha\delta
\g_{\mu\nu}-\g^{\mu\alpha}\n^\nu\delta
\g_{\mu\nu}\right)\right.\nonumber\\
&&+\frac12\n^\alpha(\varphi^2)\g^{\mu\nu}\delta
\g_{\mu\nu}-
\g^{\mu\alpha}\n_\mu\delta(\varphi^2)\label{symp2}\\
&&\left.-\frac12(\n^\alpha \phi_i\delta \phi_i-
 \phi_i\n^\alpha\delta \phi_i) -\frac12(\phi_i\n^\mu \phi_i) \g^{\nu\alpha}
\delta \g_{\mu\nu}
+\frac12g^{\mu\nu}\delta
g_{\mu\nu}(\frac12\phi_i\n^\alpha \phi_i)
\right]
\>.
\nonumber
\eea

	It can be shown by direct computation, and using the equations
of motions, that the above symplectic potential is
preserved actually.

	The conserved current associated to a diffeomorphism generated by a
vector field $X_f = f^\mu\frac{\p}{\p x^\mu}$, defined on the configuration
space of the theory as
\bea
(\delta g)_{\mu\nu}=\nabla_\mu f_\nu+\nabla_\nu f_\mu
\quad , \quad
\delta\varphi =f^\mu \p_\mu\varphi
\> ,
\label{c8}
\eea
can be written in the form:
\bea
\widehat J_f^\alpha &=&
\frac12\sqrt{-\g}
	 \left\{ \n_\mu[f^\mu\n^\alpha\varphi^2-
f^\alpha\n^\mu\varphi^2]
+
\n_\mu(\varphi^2[\n^\mu f^\alpha -\n^\alpha
f^\mu]) \right.
\nonumber \\
	& & \phantom{\frac12\sqrt{-\g}}
	\left. + \frac12 \n_\mu (f^\mu \phi_i \n^\alpha \phi_i -
	f^\alpha \phi_i \n^\mu \phi_i)
\right\}
\>.
\label{10}
\eea
It has, therefore, the form of the divergence of an antisymmetric tensor
and is, because of that, identically preserved (notice, however, that
arriving at eq. (\ref{10}) requires to use the equations of
motion).
The conserved charge associated to $\widehat{J}^\alpha_f$ can be
made explicit by noticing that the divergence of an antisymmetric
tensor $F^{\mu\nu}$ can be written, in 2D, as
\bea
\sqrt{-g} \nabla_\mu F^{\mu\nu} &=&
\sqrt{-g} \nabla_\mu \left[ \frac12 ( g^{\mu\alpha} g^{\nu\beta} -
	g^{\mu\beta} g^{\nu\alpha} ) F_{\alpha\beta} \right] =
	\nonumber \\
	&=& \sqrt{-g} \nabla_\mu \left[ \frac12 \frac{1}{g}
	\varepsilon^{\mu\nu} \varepsilon^{\alpha\beta} F_{\alpha\beta}
	\right] = \varepsilon^{\mu\nu} \partial_\nu K
\>,
\label{tresi}
\eea
with
\be
K = -\frac12 \frac{1}{\sqrt{-g}} \varepsilon^{\alpha\beta} F_{\alpha\beta}
\>.
\label{tresii}
\ee
Therefore,
\be
\widehat{J}^\alpha_f = \varepsilon^{\alpha\beta} \partial_\beta \widehat{K}
\>,
\label{tresiii}
\ee
with
\be
\widehat{K} = -\frac12 \frac{1}{\sqrt{-\widehat{g}}} \varepsilon^{\mu\nu}
	\left(
		f_\mu \widehat{\nabla}_\nu \varphi^2 +
		\varphi^2 \widehat{\nabla}_\mu f_\nu +
		\frac12 f_\mu \phi_i \widehat{\nabla}_\nu \phi_i
	\right)
\>.
\label{tresiv}
\ee

	In terms of the physical metric $g_{\mu\nu}$ the conserved
current is given by
\bea
J^\alpha_f &=& \frac12 \sqrt{-g}
	\nabla_\mu
	\left[
	\varphi^2 ( \nabla^\mu f^\alpha - \nabla^\alpha f^\mu )
	+ \frac12 ( f^\mu \phi_i \nabla^\alpha \phi_i -
		f^\alpha \phi_i \nabla^\mu \phi_i )
	\right]
	\nonumber \\
	&=& \varepsilon^{\alpha\beta} \partial_\beta K
\>,
\label{tresv}
\eea
with the charge
\be
K = -\frac12 \frac{1}{\sqrt{-g}} \varepsilon^{\mu\nu}
	\left( \varphi^2 \nabla_\mu f_\nu +
	\frac12 f_\mu \phi_i \nabla_\nu \phi_i \right)
\>.
\label{tresvi}
\ee

	It is interesting to compare (\ref{tresv}) with
Komar's formula for the conserved current in 4D \cite{Bak,Komar},
and to notice that the presence of the matter term in (\ref{tresv})
has its origin in the total divergence terms added to the
Lagrangian. On the other hand, these total divergence terms in
the Lagrangian are the reason why eq. (\ref{tresv}) differs from
other expressions for $J^\alpha_f$ given in the literature \cite{Iyer}
and, as we will see, they contribute to make $K$ finite,
under appropriate asymptotic conditions.

	From the above expressions it is not difficult to obtain,
by choosing $f^\mu=\varepsilon^{\mu\nu} x_\nu + a^\mu$ and following
the generalized Belinfante procedure \cite{Bak}, a symmetric
energy-momentum pseudotensor for the CGHS model:
\bea
\Theta^{ab}=
\frac12\p_\mu\p_\nu\left(\sqrt{-g}\varphi^2\left[\eta^{ab}g^{\mu\nu}-
\eta^{\mu b}g^{\nu a} -\eta^{\nu a}g^{\mu b}
+\eta^{\mu\nu}g^{ab}\right]\right)
\>.
\label{12}
\eea

	On the other hand, in the absence of matter, any solution
of the equations of motion can be brought, by means of a
diffeomorphism, to the form
\be
	\d s^2= -\left( \frac{m}\l - \l^2 x^+ x^- \right)^{-1}
		\d x^+\d x^-
		\quad,\quad
	\varphi^2=\frac{m}\l -\l^2 x^+x^-
\> ,
\label{solkruskal}
\ee
where $x^+, x^-$ can be considered as the null Kruskal coordinates.
The spacetime has four regions which can be characterised by the sign
of the Kruskal coordinates. The asymptotic flat regions are
characterised by $-\l^2x^+x^->0$. In the region $I$
($x^+>0$, $x^-<0$) the metric can be written in a static
asymptotically-flat form:
\bea
\d s^2&=& -\left(1+\frac{m}\lambda\e^{-2\lambda
\sigma}\right)^{-1}\d \sigma^+\d\sigma^-
\label{metricaflat}
\>,\\
\e^{-2\Phi}&=&\frac{m}\lambda+\e^{2\lambda\sigma}
\label{vpflat}
\>,
\eea
by means of the coordinate change
\bea
\lambda x^+ &=& e^{\lambda(\tau+\sigma)} \>, \label{cambioi}\\
\lambda x^- &=& -e^{-\lambda(\tau-\sigma)} \> . \label{cambioii}
\eea

	In the other asymptotically flat region $II$ ($x^+<0\>,\>x^->0$)
the static metric can be achieved by the change
\bea
\lambda x^+ &=& - e^{\lambda(\tau+\sigma)} \>, \label{cambioiii}\\
\lambda x^- &=& e^{-\lambda(\tau-\sigma)} \> . \label{cambioiv}
\eea

	If we  calculate the energy of the
basic solution of the CGHS model, eq. (\ref{metricaflat},\ref{vpflat}),
by means of this E-M pseudotensor
we will, surprisingly,
not find any sensible result. In fact, the resulting expression
is divergent and even do not involve the constant $m$.
In the next sections, we will find the explanation for this result:
the construction of a symmetrized E-M
pseudotensor requires the theory to be invariant
under asymptotic Lorentz transformations. We will show, however,
that in order to  have a well defined physical theory, we can not allow
pure-Lorentz asymptotic rotations.

	Going back to eq. (\ref{tresv}),  the contribution
to the conserved charge for the
basic solution in (\ref{metricaflat}-\ref{vpflat}) is:
\bea
K_I
&=&\frac12\left\{
\left({m\over\lambda}+e^{2\lambda\sigma}\right) \p_\sigma f^\tau +
\left({m\over\lambda}+e^{2\lambda\sigma}\right) \p_\tau f^\sigma +
2mf^\tau \right\}_{|_{\sigma\rightarrow+\infty}}
\>,
\label{c30}
\eea
where the subindex $I$ refers to the region in which the above current
has been evaluated. With the asymptotic fall-off conditions
\bea
e^{2\lambda\sigma} \partial_\sigma f^\tau
	&\stackrel{\sigma\rightarrow\infty}{\sim}& 0 \>, \nonumber \\
e^{2\lambda\sigma} \partial_\tau f^\sigma
	&\stackrel{\sigma\rightarrow\infty}{\sim}& 0 \>,
\label{c30i}
\eea
the Noether charge associated with the Killing time translation
($f^\tau \stackrel{\sigma\rightarrow\infty}{\sim} 1$)
is $K_I = m$.
Terms like $\lambda e^{2\lambda\sigma} f^\tau$,
that would appear in the expression for the Noether charge had we
started with Lagrangian (\ref{CGHS2}), cancel out in (\ref{c30}).
It is just the Lagrangian (\ref{CGHS4}) which gives directly the
finite terms only. The reason is that the Noether charge (\ref{c30})
can be seen as the result of contracting the presymplectic potential
with the infinitesimal diffeomorphism $X$ associated with
the asymptotic time translation (in region I). Both quantities are
well defined in the covariant phase space, as we will see in the
next section. The charges associated with the asymptotic spatial
translations and Lorentz transformations
are zero and divergent, respectively.

	Moreover we also want to stress that the Noether charge
(\ref{c30}) just gives the mass of the black hole without the
discrepant factor $\frac12$, as happens in the Komar's formula
for energy in General Relativity. We shall also understand
this fact in the context of the canonical formalism.

\section{Canonical structure and asymptotic symmetries
	of the CGHS model.}
\label{E}

	Let us begin our analysis of the canonical structure
of the CGHS model by writing the general classical solution
of the theory without matter. It is well known that any solution
is equivalent under diffeomorphisms to the solution
\be
\d \hat s^2 = -\d x^+\d x^-
	\quad,\quad
\varphi^2=\frac{m}\l -\l^2 x^+x^-
\>.
\label{E5}
\ee
The solutions are characterized by an unique diffeomorphism
invariant parameter, $m$, and therefore the variable
canonically conjugate to $m$ should be ``hidden''
in the group of diffeomorphisms.
The situation is somewhat similar to the trivial example
of the free particle. Any solution is equivalent, under the
Galileo group, to the one with the particle lying at rest and,
therefore, the canonical degrees of freedom of the system are
found in the symmetry (Galileo) group.

	Our aim now is to find the degrees of freedom of the theory
that are ``hidden'' in the group of diffeomorphisms.
To this end we shall compute explicitly the two-form (\ref{dosii})
(more precisely, the potential one-form (\ref{dosiii}) ). This
requires to adjust the boundary condition adequately for the
potential form to be finite and independent of the spacelike Cauchy
surface. Therefore, we shall assume the metric $g_{\mu\nu}$ to be
flat at spatial infinity with a specific fall-off behaviour.

	Let us apply a general diffeomorphism to the basic solution
(\ref{E5}). We find
\bea
\d \hat s^2 &=& -\d P\d M \>, 	\label{E9}\\
\varphi^2 &=& \frac{m}\l-\l^2 PM \>, \label{E10}
\eea
where $P$ and $M$ are two arbitrary functions
$P,M\>:{\cal M}\rightarrow \R$;
$x^+ = P(\tau,\sigma)$, $x^- = M(\tau,\sigma)$.

	We have
\bea
\g_{\mu\nu} &=& -\frac12\left(\p_\mu P\p_\nu M + \p_\mu M \p_\nu
P\right)
\>,
\label{E11}\\
\sqrt{-\g} &=& \frac12\varepsilon^{\alpha\beta}\p_\alpha P\p_\beta M
\>,
\label{E12}\\
\g^{\mu\nu} &=&
-\frac1{(\sqrt{-\g})^2}\varepsilon^{\mu\alpha}
	\varepsilon^{\nu\beta}\g_{\alpha\beta}
\>,
\label{E13}\\
\hG_{\mu\alpha\beta} &=&
	-\frac12\left[\p_\alpha \p_\beta P \p_\mu M +
	\p_\alpha \p_\beta M \p_\mu P\right]
\>,
\label{E14}\\
\hG^\mu_{\alpha\beta} &=& \frac1{2\sqrt{-\g}}
\varepsilon^{\mu\nu}
\left[\p_\alpha \p_\beta P \p_\nu M - \p_\alpha \p_\beta M \p_\nu P\right]
\>.
\label{E15}
\eea
Obviously we also have
\be
\n_\alpha \n_\beta M = 0
	\quad,\quad
\n_\alpha \n_\beta P = 0
	\quad,\quad
\forall a,b
\>,
\label{E16}\ee
and, therefore, for the metric parametrized as in eq. (\ref{E11}):
\bea
\delta \g_{\alpha\beta}
	&=& -\frac12\left(\p_\beta M\delta \p_\alpha P
+\p_\alpha P\delta\p_\beta M \right.\nonumber\\
	&&\left.+\p_\beta P\delta \p_\alpha M
+\p_\alpha M\delta\p_\beta P\right)\nonumber\\
	&=& -\frac12\left[\n_\alpha(\n_\beta M\delta P +\n_\beta P\delta M)+
\n_\beta(\n_\alpha P\delta M +\n_\alpha M\delta P)\right]\label{E17}\\
	&=& \n_\alpha h_\beta +\n_\beta h_\alpha
\>,
\eea
where the one-form $h_\mu$ is given by:
\be
h_\mu = -\frac12\left(\n_\mu P\delta M + \n_\mu M\delta
P\right)
\quad\Rightarrow\quad
h^\alpha = -\frac12\g^{\alpha\mu}\left(\n_\mu P\delta M +
\n_\mu M\delta P\right)
\>.
\label{E18}
\ee
We can easily see that, with the one-form $h^\alpha$ defined
above, we can write as well:
\be
\n_\mu \delta \vp^2 = \n_\mu (h^\alpha \n_\alpha\vp^2)
\quad,\quad \forall \mu
\>.
\label{E18b}
\ee
So, to get the symplectic potential for the general solution
given in  (\ref{E9}-\ref{E10}), it is enough to replace in eq.
(\ref{10}) the diffeomorphism $f^\mu$ by the quantities $h^\mu$ as
defined in (\ref{E18}).

	The symplectic potential will therefore be given by the divergence
of an antisymmetric tensor ($K$ is now a one-form)
\be
\widehat{j}^\alpha = \varepsilon^{\alpha\mu} \partial_\mu K
\>,
\label{E19}
\ee
and the symplectic form will be a pure-boundary term, thus
implying that the theory has a finite number of degrees of
freedom.

\subsection{Conditions of flatness at spatial infinity.}
\label{f}

	The condition for the metric to be flat at spacelike infinity
means:
\be
g_{\mu\nu} = -\frac12\frac{\p_\mu P\p_\nu M + \p_\mu M\p_\nu P}
{\bar{m} - \lambda^2 PM}
\spacelike
\eta_{\mu\nu}
\>,
\label{f1}
\ee
where $\eta_{\mu\nu}=\left( \begin{array}{cc} -1 & 0 \\
					      0 & 1 \end{array} \right)$
and $\bar{m} = \frac{m}{\lambda}$.
%

	In region I ($P \equiv x^+ > 0$ and $M \equiv x^- < 0$)
we can make
\be
\lambda P = \e^{\lambda C}
\quad,\quad
-\lambda M = \e^{\lambda R}
\>.
\label{f5}
\ee
Using (\ref{f1}) and because of $-PM \spacelike +\infty$,
we arrive at
\be
\d C\>\d R \spacelike -\d\tau^2 + \d\sigma^2
\>,
\label{f6}
\ee
or, what is the same,
\bea
&& \dot C\dot R \spacelike -1 \>, \nonumber\\
&& \dot C\R' +C'\dot R \spacelike 0 \>, \\
&& C'R'\spacelike 1 \>, \nonumber
\eea
requirements whose solution can be written in the form:
\bea
C(\tau,\sigma) &=&
	\alpha \sigma^+ + A + U(\tau,\sigma)  \>,	\label{f7}\\
R(\tau,\sigma) &=&
	-\frac1\alpha \sigma^- - B + V(\tau,\sigma) \>,	\label{f8}
\eea
where $\alpha$, $A$ and $B$ are real numbers,
$\sigma^+ = \tau + \sigma$, $\sigma^- = \tau - \sigma$, and
\be
U,V \spacelike 0
\>.
\label{f9}
\ee

	The interpretation of (\ref{f7}-\ref{f8}) on the light of (\ref{f6})
is obvious: the only allowed diffeomorphisms $(C,R)$ are those that
asymptotically are Poincar\'e transformations in the coordinates $\tau$,
$\sigma$. Surprisingly we will find additional constraints on the
asymptotic transformations in the computation of the (on-shell) symplectic
potential.

\subsection{Symplectic potential.}

	The symplectic current potential is given by
\be
\widehat{j}^\alpha = \varepsilon^{\alpha\mu} \partial_\mu \widehat K
\>,
\label{cuatrodosi}
\ee
with $\widehat K$, formally, given by (\ref{tresiv}).

	The first consequence of the above formulas is that
the symplectic potential reads as
\be
\theta = \int_\Sigma \p_\mu \widehat{K} \d x^\mu
	= \widehat{K} (i^0_R) - \widehat{K} (i^0_L)
\>,
\label{f13i}
\ee
where $\Sigma$ is an arbitrary Cauchy surface (see Fig. I).
We have to stress that $\Sigma$ is not required to intersect
the bifurcation point of the horizon as it was in Ref. \cite{Navarro}.
The point now is to show that the one-form $K$ can have well defined
values in the right and left spatial infinities. In fact we
shall find that not all the asymptotic Poincar\'e transformations
are permitted in order to have a well defined result for $\theta$ (i. e.,
independent of the Cauchy surface).

\ifnum\opcion=1
	\begin{figure}
	\centerline{\psfig{figure=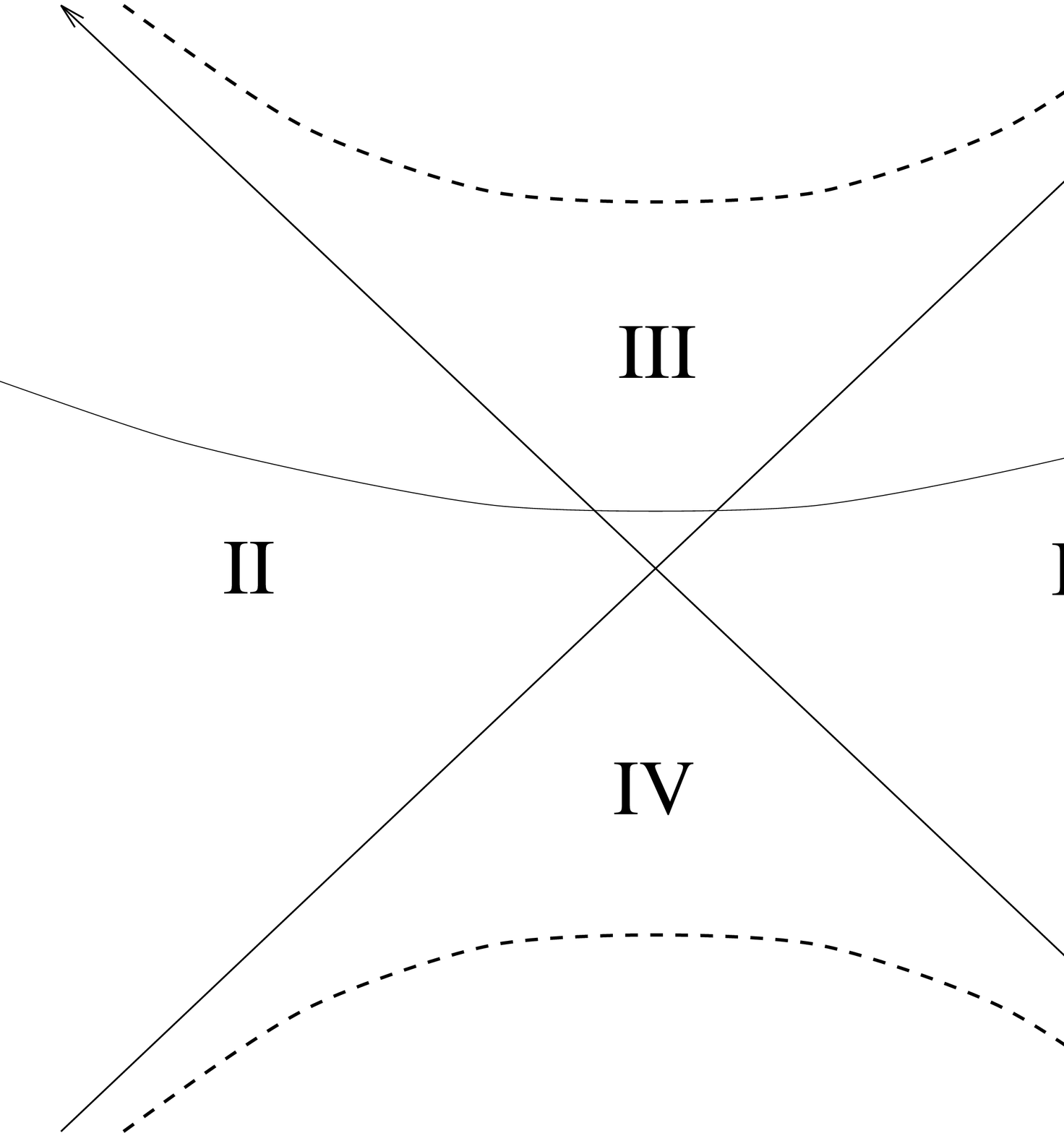,width=4in}}
	\caption{Kruskal diagram for a black hole spacetime. $\Sigma$
		is an arbitrary Cauchy surface.}
	\end{figure}
\else
\fi

	Replacing the ``diffeomorphism'' $h^\mu$ by its expression in eq.
(\ref{E18}) we find after a bit of algebra:
\bea
\widehat{K}(P,M,m) &=&
	-\frac12 \lambda^2 \left(P\delta M - M\delta P\right)
	\nonumber\\
	&& -\frac1{4\sqrt{-\g}}
	\lambda^2 PM\varepsilon^{\l\rho}\left(\p_\rho P\delta\p_\l M
	+\p_\rho M\delta \p_\l P\right)
	\\
	&& +\frac{\bar{m}}{4\sqrt{-\g}}
	\varepsilon^{\l\rho}\left(\p_\rho P\delta\p_\l M
	+\p_\rho M\delta \p_\l P\right)
	\nonumber
\>,
\eea
and, after having made use of (\ref{f5}), we find
\bea
\widehat{K} (C,R,m) &=&
	-\frac1{4\chi} \lambda^2 PM\varepsilon^{\l\rho}
  	\left(\p_\rho C\delta\p_\l R + \p_\rho R\delta \p_\l C\right)
	\nonumber\\
 	&& +\frac{\bar{m}}{4\chi}\varepsilon^{\l\rho}
  	\left(\p_\rho C\delta\p_\l R
  	+\p_\rho R\delta \p_\l C\right)
	\\
 	&& -\frac{m}2\delta(R-C)
	\nonumber
\>,
\eea
where $\chi$ is given by:
\be
\chi = \frac12\varepsilon^{\l\rho}\p_\l C\p_\rho R \spacelike 1
\> .
\ee
And after the replacements in (\ref{f7},\ref{f8}) we find
\bea
\widehat{K} &=&
	\frac1{4\chi}(\bar{m} - \lambda^2 PM)
	\varepsilon^{\l\rho}\left(\p_\rho U\delta\p_\l V
	+\p_\rho V\delta \p_\l U\right) \label{f14}
	\nonumber \\
 && + \frac1{2\chi}(\bar{m} - \lambda^2 PM)
	(\frac{1}{\alpha} \delta \pp U + \alpha \delta \mm V)
	\nonumber \\
 && - \frac1{2\chi} (\bar{m} - \lambda^2 PM)
	(\partial_+ U \> \delta \frac{1}{\alpha} +
	\partial_- V \> \delta\alpha)
	\nonumber\\
 && - \frac{m}{2} \> \delta (V-U)
	\\
 && + \frac1{2\chi}(\bar{m} - \lambda^2 PM)
	\frac2\alpha\delta\alpha
	\label{f16}
	\nonumber\\
 && + \frac{m}2\delta(\frac1\alpha \sigma^- + \alpha \sigma^+)
	\label{f17}
	\nonumber \\
 && + m \> \delta\left(\frac{A+B}{2}\right)
	\nonumber
\>.
\label{f18}
\eea

	It is easy to realize from the last expression above that for
the symplectic potential to be finite and independent of the spacelike
Cauchy surface (i.e, independent of $\tau$), requires first that
$\alpha=1$. That is to say, the Lorentz transformations are not
allowed. So that, we are left with:
\bea
\widehat{K} &=&
	\frac1{4\chi}(\bar{m} - \lambda^2 PM)
	\varepsilon^{\l\rho}\left(\p_\rho U\delta\p_\l V
	+\p_\rho V\delta \p_\l U\right)
	 \nonumber \\
&& 	+ \frac1{2\chi}(\bar{m} - \lambda^2 PM)
	(\delta \pp U + \delta \mm V)
	\label{f20} \\
&&	+ m \> \delta\left(\frac{A+B}{2}\right)
	 \nonumber
\>.
\eea
Moreover, to find a finite resulting expression for (\ref{f20})
we have to require an appropriate asymptotic fall-off for the functions
$U$ and $V$.
 From a close inspection of eq. (\ref{f20}), and taking into
account the asymptotic behaviour of $-PM$, it is not difficult
to realize that the most natural requirement in order to have
a sensible reduced phase space is
\bea
e^{2\lambda\sigma} \dot{U}, e^{2\lambda\sigma} \dot{V}
	&\stackrel{\sigma\rightarrow\infty}{\sim}& 0
\>, \nonumber \\
e^{2\lambda\sigma} U', e^{2\lambda\sigma} V'
	&\stackrel{\sigma\rightarrow\infty}{\sim}& 0
\>.
\label{f21ii}
\eea
Therefore we have arrived at:
\be
\widehat{K} (i^0_R) = m \> \delta\left( \frac{A+B}{2} \right)
\>,
\label{f21iii}
\ee
where $\frac{A+B}{2} \equiv f(i^0_R)$ is the Killing time translation
at right spatial infinity.

	In the other asymptotically flat region $x^+ = P(\tau,\sigma) < 0$,
$x^- = M(\tau,\sigma) > 0$, we should write
\be
-\lambda P = e^{\lambda C} \quad,\quad \lambda M = e^{\lambda R}
\label{f21iv}
\ee
instead of (\ref{f5}), where the asymptotic flatness requires that
(the asymptotic Lorentz transformation has already been neglected)
\bea
C(\tau,\sigma) &=& \tau + \sigma + A + U(\tau,\sigma) \>, \nonumber \\
R(\tau,\sigma) &=& -(\tau - \sigma) - B + V(\tau,\sigma) \>, \\
  & & U, V \stackrel{\sigma\rightarrow\infty}{\sim} 0 \>.
\label{f21v}
\eea

	Proceeding in the same way as in the region I we obtain
\be
\widehat{K} (i^0_L) = m \> \delta\left( \frac{A+B}{2} \right) \>,
\label{f21vi}
\ee
where now $\frac{A+B}{2} \equiv f(i^0_L)$ stands for the Killing
time translation at left infinity.

	Taking into account (\ref{f21iii}) and (\ref{f21vi}) we obtain
the final expression for the symplectic potential
\be
\theta = m \> \delta\left( f(i^0_R) - f(i^0_L) \right)
\>.
\label{f21vii}
\ee

\subsection{Diffeomorphisms in the presence of matter.}
\label{g}

	When matter is present, the procedure applied above is much more
complicated. This is so because we would not be able to
write the symplectic form as  a pure boundary
term. The model has an infinite number
of degrees of freedom and, because of that, the symplectic form has,
unavoidably, a bulk term. Intuitively we expect, however,
that diffeomorphisms should be ``almost'' pure gauge.
In the covariant formalism, this means that the presymplectic
two-form (\ref{dosii}) should be degenerated along the directions
that corresponds to the gauge transformations of the theory.
We can arrive at this result by contracting the symplectic two-form
with the generator of a diffeomorphism:
\bea
(\delta g)_{\mu\nu}=\nabla_\mu f_\nu+\nabla_\nu f_\mu
\quad,\quad
\delta\varphi =f^\mu \p_\mu\varphi
\quad,\quad
\delta \phi_i=f^\mu\p_\mu \phi_i
\>.
\label{g1}
\eea

	The only linearized equation of motion which is not trivial
to obtain is:
\be
\n_\mu \delta\hG^\mu_{\alpha\beta} - \n_\alpha \delta\hG^\mu_{\beta\mu}=0
\>.
\label{g3}
\ee
and, after a long computation, we arrive at:
\be
\hi_{X_f} \> \delta j^\alpha = \partial_\lambda T^{\lambda\alpha}
\>,
\ee
i.e. $\hi_{X_f} \omega$ is a pure boundary term, with
\bea
T^{\lambda\alpha} = -T^{\alpha\lambda} &=&
\frac12\sqrt{-\g}
\left\{
\vp^2\left[-\delta\log\sqrt{-\g}(\n^\lambda
f^\alpha - \n^\alpha f^\lambda)\right.
\right.
\nonumber\\
&&\qquad+(\delta g^{\mu\alpha}\n_\mu f^\lambda -\delta
g^{\mu\lambda}\n_\mu f^\alpha)\nonumber\\
&&\qquad+(f^\lambda \g^{\mu\nu}\delta\hG^\alpha_{\mu\nu}-
f^\alpha g^{\mu\nu}\delta\hG^\lambda_{\mu\nu})\nonumber\\
&&\qquad+(f^\nu\g^{\mu\alpha}\delta\hG^\lambda_{\mu\nu}-
f^\nu g^{\mu\lambda}\delta\hG^\alpha_{\mu\nu})\nonumber\\
&&\left.\qquad+(f^\alpha\g^{\mu\lambda}\delta\hG^\nu_{\mu\nu}-
f^\lambda g^{\mu\alpha}\delta\hG^\nu_{\mu\nu})\right]\label{g4}\\
&&-\delta\vp^2(\n^\lambda
f^\alpha - \n^\alpha f^\lambda)\nonumber\\
&&+2(f^\alpha\n^\lambda\delta\vp^2- f^\lambda\n^\alpha\delta\vp^2)\nonumber\\
&&+(f^\alpha\n_\mu\vp^2\delta g^{\lambda\mu}
-f^\lambda\n_\mu\vp^2\delta g^{\alpha\mu})\nonumber\\
&&\left.
+(f^\alpha\n^\lambda\phi_i\delta \phi_i -
f^\lambda\n^\alpha\phi_i\delta\phi_i)
\right\}
\nonumber
\>.
\eea

	It is convenient now to rewrite the expressions above
in terms of the physical metric, which has a better behaviour at
spacelike infinity. We find:
\bea
\frac2{\sqrt{-g}}T^{\lambda \alpha}&=&
\vp^2\left\{-\delta\log{\sqrt{-g}}\left(\nabla^\lambda f^\alpha
-\nabla^\alpha f^\lambda\right)\right.\nonumber\\
{}
&&\qquad-\delta\log{\sqrt{-g}}\left(f^\alpha\nabla^\lambda \log\vp^2-
f^\lambda\nabla^\alpha\log\vp^2\right)\nonumber\\
{}
&&\qquad+\left(\delta g^{\mu\alpha}\nabla_\mu f^\lambda -
\delta g^{\mu\lambda}\nabla_\mu f^\alpha\right)\nonumber\\
{}
&&\qquad-\frac12\left(\delta g^{\mu\alpha}\nabla_\mu\log\vp^2 f^\lambda -
\delta g^{\mu\lambda}\nabla_\mu \log \vp^2f^\alpha\right)\label{g8}\\
{}
&&\qquad+\frac12\left(\nabla^\alpha\log\vp^2 f_\mu\delta g^{\mu\lambda}
-\nabla^\lambda\log\vp^2 f_\mu\delta g^{\mu\alpha}\right)\nonumber\\
{}
&&\qquad-\delta\log{\vp^2}\left(\nabla^\lambda f^\alpha
-\nabla^\alpha f^\lambda\right)\nonumber\\
{}
&&\qquad\left.+2\left(f^\alpha\nabla^\lambda\delta\log\vp^2-
f^\lambda\nabla^\alpha\delta\log\vp^2\right)\right\}\nonumber\\
{}
&&+\left(f^\alpha\nabla^\lambda\phi_i\delta\phi_i-
f^\lambda\nabla^\alpha\phi_i\delta\phi_i\right)\nonumber
\>.
\eea

	If we take, as it appears the most natural,
boundary conditions such that
$\phi_i\spacelike0$, the expressions above indicates clearly that the
analysis of the model with matter reduces itself to the case without
matter.
Therefore, the contribution of diffeomorphisms to the reduced
phase space of the theory is the same when there are matter
fields as when there are not.
For instance, if we take into account that
$\vp^2\sspacelike\e^{2\lambda\sigma}$, we see that the leading term in
(\ref{g8}) behaves as $-\delta\log{\vp^2}\vp^2\left(\nabla^\lambda f^\alpha
-\nabla^\alpha f^\lambda\right)$. The finiteness of this term implies
$\varepsilon^{\mu\nu}\p_\mu f_\nu\spacelike0$, thus forbidding as symmetries
of the theory those diffeomorphisms that are asymptotically
Lorentz transformations.

\section{Symplectic potential of Schwarzschild black holes.}

	The symplectic current potential of general relativity in
vacuum is given by
\be
j^\alpha = \frac{1}{16\pi} \sqrt{-g}
	\left( g^{\mu\nu} \delta\Gamma^{\alpha}_{\mu\nu} -
		g^{\mu\alpha} \delta\Gamma^\nu_{\mu\nu} \right)
\>,
\label{cincoi}
\ee
and due to the Hilbert-Einstein Lagrangian vanishes on-shell the
current (\ref{cincoi}) is conserved. In this section we shall
work out the symplectic potential associated with the Schwarzschild
black hole solutions. Instead of starting with the basic solution
and acting on it with a general diffeomorphism we shall assume that
the relevant asymptotic symmetry is the Killing time translation.
Therefore we can write the general solution in regions I and II
as follows
\be
ds^2 = -\left( 1-\frac{2m}{r} \right) d(t+f(t,r))^2 +
	\left( 1-\frac{2m}{r} \right)^{-1} dr^2 +
	r^2 \left( d\theta^2 + \sin^2 \theta d\varphi^2 \right)
\>.
\label{cincoii}
\ee
In addition, we shall choose the Cauchy surface in such a way that
it connects the spatial infinities through the asymptotically
flat regions I and II.

	The symplectic potential is the integral of an exact
three-form and, therefore, it receives contribution from the
two-spheres $S^2_{R,L}$ at infinity only
\bea
\theta &=& \frac{1}{16\pi} \int_{S^2_R} d\Omega \> \sin\theta
	\left[ -\frac{r(r-2m)}{1+\dot{f}} f' \>\delta f +
	r(r-2m) \>\delta f' \right.
	\nonumber \\
	&& \left.\phantom{\frac{1}{16\pi} \int_{S^2_R} d\Omega \> \sin\theta [}
	- 2 f \>\delta m + 2 m \>\delta f \right]
	\nonumber \\
&& -\frac{1}{16\pi} \int_{S^2_L} d\Omega \sin\theta
	\left[ -\frac{r(r-2m)}{1+\dot{f}} f' \>\delta f +
        r(r-2m) \>\delta f' \right.
	\nonumber \\
	&& \left.\phantom{-\frac{1}{16\pi} \int_{S^2_R} d\Omega \> \sin\theta [}
	- 2 f \>\delta m + 2 m \>\delta f \right]
	\label{cincoiii}
\>.
\eea

	To obtain a well-defined result we have to assume the
following fall-off behaviour:
\be
r^2 f',  \dot{f} \stackrel{r\rightarrow\infty}{\sim} 0
\>.
\label{cincoiv}
\ee
With the prescribed fall-off the integral (\ref{cincoiii}) turns out
to be
\be
\theta = \frac12
	\left[
		m \> \delta \left( f(i^0_R) - f(i^0_L) \right) -
		\left( f(i^0_R) - f(i^0_L) \right) \> \delta m
	\right]
\>.
\label{cincovi}
\ee

\section{Conclusions and final comments.}

	On the light of the result of Secs. 3-4 we observe that
the asymptotic fall-off behaviour of the diffeomorphisms entering
in the symplectic potential (\ref{c30i}) are similar to
that required to have a well-defined Noether charge (\ref{f21ii}).
This is a consequence of the closed relationship between
the canonical formalism and the Noether theorem outlined in
Sec. 2.

	Using the covariant phase space picture we have
determined the canonical structure of the CGHS model
in the absence of matter, and the character of the
asymptotic symmetries, without any a priori assumption on the
dilaton asymptotic behaviour. The requirements made in 4.1-2
on the metric are enough to arrive at a clear result.
The difference of Killing
time translations at spatial infinities turns out to be
the conjugate variable to the black hole mass.
The asymptotic spatial translations are ``gauge''-type symmetries:
they decouple in the symplectic potential and leads to trivial
Noether charge. The asymptotic Lorentz transformation breaks down
(it cannot be permitted to have a well-defined symplectic form)
and leads to a divergent Noether charge.
This results are closely related. On general grounds, the action
of a Lorentz transformation gives linear momentum to the system.
In the CGHS model it breaks down and, therefore, the linear momentum
vanishes identically, in accord with the ``gauge'' nature
of the spatial translations for the model.
This provides an
explanation for the failure of the symmetric energy-momentum
pseudotensor. The definition of this quantity requires the
theory to be invariant under asymptotic Lorentz transformations
and we have shown that this is not the case for the CGHS model.

	As a byproduct of our study we also provide an explanation
of the well-known factor 2 in the Komar formula for the mass in
General Relativity. Although both the stringy and Schwarzschild
black holes have the same symplectic structure
\be
\omega = \delta\theta =
	\delta m \wedge \delta \left( f(i^0_R) - f(i^0_L) \right)
\>,
\label{seisi}
\ee
they differ in the form of the symplectic potential.
For the CGHS black hole the potential contains only the term
with $\delta\left( f(i^0_R) - f(i^0_L) \right)$
\be
\omega_{CGHS} = m \> \delta\left( f(i^0_R) - f(i^0_L) \right)
\>.
\label{seisii}
\ee
The corresponding Noether charge associated with a (right)
asymptotically Killing time translation is just the black
hole mass
\be
\theta_{CGHS} \left( \frac{\partial}{\partial f(i^0_R)} \right) = m
\>.
\label{seisiii}
\ee

	In the case of Schwarzschild black hole the symplectic
potential contains a term with $\delta m$ as well. So that the
Noether charge cannot coincide exactly with the mass.
Since the potential is symmetric in $m$ and $f(i^0_R) - f(i^0_L)$
\be
\theta_{Sch} = \frac12
	\left(
	m \> \delta\left( f(i^0_R) - f(i^0_L) \right) -
	\left( f(i^0_R) - f(i^0_L) \right) \> \delta m
	\right)
\>,
\label{seisiv}
\ee
the Noether charge is actually one half of the mass
\be
\theta_{Sch} \left( \frac{\partial}{\partial f(i^0_R)} \right) =
	\frac{m}{2}
\>.
\label{seisv}
\ee

\section*{Acknowledgments}

	M. N. would like to thanks L. J. Garay for valuable discussions.
M. N. acknowledges the MEC for financial support.
C. F. T. is grateful to the Generalitat Valenciana for a FPI
grant. This work was partially supported by the CICYT and
the DGICYT.

\ifnum\opcion=1
\else
	\setlength{\anchopar}{\hsize-\rightskip-\leftskip}
	\newpage

	\pagestyle{empty}
	\section*{Figure Captions}
	Figure I: Kruskal diagram for black hole spacetime. $\Sigma$
	is an arbitrary Cauchy surface.

	\newpage
	\pagestyle{empty}
	\centerline{\Large FIGURE I}
	\vfill
	\begin{figure}[h]
	\centerline{\psfig{figure=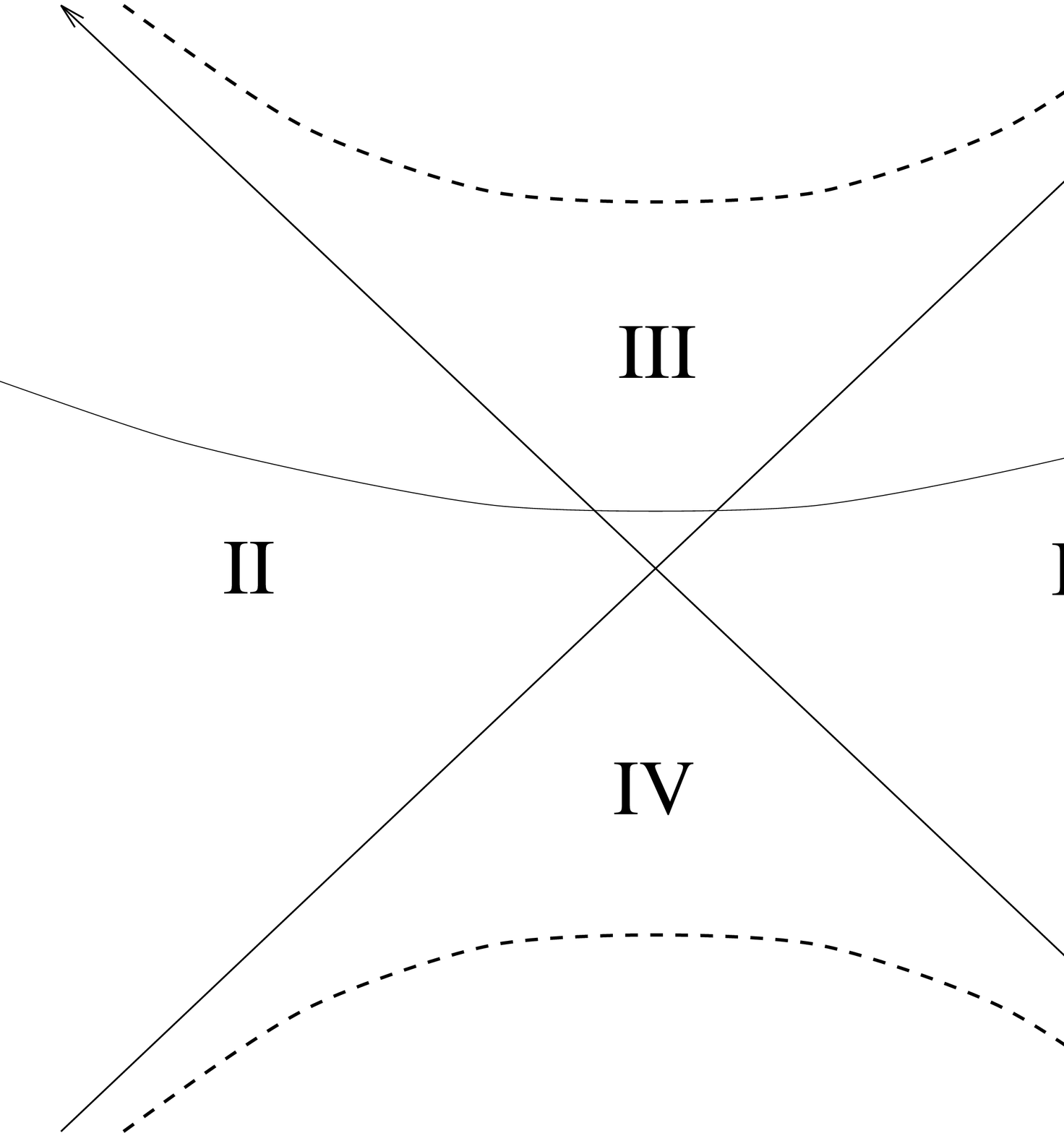,width=\anchopar}}
	\end{figure}
	\vfill
\fi

\end{document}